**Benchmarking Active Learning Strategies for Materials Optimization and Discovery**

Alex Wang[1] (0000-0002-4235-0477), Haotong Liang[1] (0000-0002-3119-013X), Austin McDannald[2] (0000-0002-3767-926X), Ichiro Takeuchi[1] (0000-0003-2625-0553), A. Gilad Kusne[1,2] (0000-0001-8904-2087)

1. Materials Science & Engineering Dept, University of Maryland, College Park MD
2. Materials Measurement Science Division, National Institute of Standards and Technology, Gaithersburg MD

**ABSTRACT**

Autonomous physical science is revolutionizing materials science. In these systems, machine learning controls experiment design, execution, and analysis in a closed loop. Active learning, the machine learning field of optimal experiment design, selects each subsequent experiment to maximize knowledge toward the user goal. Autonomous system performance can be further improved with implementation of scientific machine learning, also known as inductive bias-engineered artificial intelligence, which folds prior knowledge of physical laws (e.g., Gibbs phase rule) into the algorithm. As the number, diversity, and uses for active learning strategies grow, there is an associated growing necessity for real-world reference datasets to benchmark strategies. We present a reference dataset and demonstrate its use to benchmark active learning strategies in the form of various acquisition functions. Active learning strategies are used to rapidly identify materials with optimal physical properties within a ternary materials system. The data is from an actual Fe-Co-Ni thin-film library and includes previously acquired experimental data for materials compositions, X-ray diffraction patterns, and two functional properties of magnetic coercivity and the Kerr rotation. Popular active learning methods along with a recent scientific active learning method are benchmarked for their materials optimization performance. We discuss the relationship between algorithm performance, materials search space complexity, and the incorporation of prior knowledge.

**INTRODUCTION**

Technological advances are often dictated and driven by materials discovery. The need for ever-better materials spurs modern scientists to explore materials of greater and greater complexity. For instance, interest in high-temperature superconductors has grown from the study of single-element materials to complex compounds such as Hg-Tl-Ba-Ca-Cu-O[1]. Similarly, the number of elements used in the electronics industry increased from around ten to fifty during the 1990s[2]. However, with each newly-added stoichiometric element or processing parameter, the number of possible materials to investigate grows exponentially. As a result, the traditional expert-driven Edisonian, one-by-one trial-and-error approach is rapidly becoming impractical.

In these Edisonian studies, materials scientists first select a target materials system (e.g. a ternary system A-B-C consisting of chemical elements A, B, and C) to investigate, bounding the study to a composition and processing space. This in turn determines experiment setup, such as the fabrication method, the materials sources to use, and appropriate materials processing equipment. Prior knowledge is then used to build a model, heuristic, or intuition to predict desired material properties from materials synthesis parameters. The materials scientist then uses the predictive model to guide subsequent materials synthesis, characterization, and analysis.

Combinatorial high-throughput (CHT) strategies were developed to enhance the rate and the efficiency of materials exploration[3]. CHT strategies allow for hundreds to thousands of materials from a target materials system (A-B-C) to be synthesized in parallel as a composition library. The library consisting of hundreds of different compositions $A_xB_yC_{1-x-y}$ (where, x and y are the compositional parameters varied on the library wafer layout with increment, for instance, of 0.01 (with $0 \leq 1-x-y \leq 1$)) is then loaded into a characterization system where each material is measured in rapid succession. However, for characterization methods of high cost or time, such as X-ray photoelectron spectroscopy or determination of the band gap, measuring all of the hundreds (or even sometimes thousands) of samples within a given library can be prohibitive. This challenge motivated the use of active learning, or the machine learning field of optimal experiment design, to guide sequence of measurement experiments across the library[4]. Each measurement is selected to maximize knowledge toward a user specific goal, e.g., identifying the compositional parameter x and y which gives the optimal physical property. An important associated goal is often to determine the composition-phase map across the entire ternary A-B-C. The use of active learning enables a more streamlined procedure for screening the materials phase space and provides the ability to do on-the-fly, adaptive, and iterative learning and optimization with a minimal number of experiments[5]. Depending on the complexity of the composition-property landscape across the ternary, it is possible to arrive at the "correct answer," i.e. the optimum composition after only a fraction of the entire library is measured. This aspect of active learning is particularly attractive when the search parameter space is extended to multiple dimensions beyond mapping of ternary phase diagrams.

The use of active learning in materials science gained popularity half a decade ago, with active learning driving recommendation engines to guide experimentalists in the lab[6]. Active learning is often combined with a machine learning (ML) surrogate model for when the underlying mechanistic model is unknown. Such active learning tools provided improved performance over traditional optimal experiment design (OED) methods[7], as the predictive model (i.e., "response model") is updated upon each iteration, ensuring always optimized decision making[8]. More recently active learning has been integrated into autonomous materials research systems capable of performing experiment design, execution, and analysis in a closed-loop[9]. For example, autonomous systems have been used to optimize materials processing parameters to tune quantum dots optical behavior[10], or come up with best molecular mixtures for improved photovoltaics films[11], and identify the new best-in-class phase change memory material – the first autonomous discovery of a best-in-class material[4]. The success of these autonomous systems depends on the active learning schemes employed.

Each active learning scheme pairs a predictive model, used to "forecast" the properties of yet-to-be-measured materials with an acquisition function, which defines the utility of investigating each possible material. Materials compositions of maximal utility are then selected for subsequent studies. A probabilistic predictive model provides added advantage: these models output both an estimate and uncertainty for predictions, an example being the Gaussian Process[12] (GP) which is used here. Prediction estimate and uncertainty can then be combined in a Bayesian optimization (BO) algorithm to define utility[13]. When the task is materials optimization, successive BO acquisition functions strike a balance between an exploratory methods and exploitative methods. Exploratory methods seek global knowledge of the unknown target function, exemplified by choosing materials where the prediction model has maximum uncertainty. Exploitation methods search for optima of the target function, exemplified by choosing materials with predicted property optima. The combination allows an active learning scheme to avoid falling into local optima and provides greater speed and stability in the search for global optima[13]. The performance of each active learning scheme is dependent on the selection of the materials model, the acquisition function, and the materials challenge and system being addressed.

As the number of active learning schemes increases, and the target materials challenges become more complex, the choice of active learning scheme becomes more important. The wrong selection can lead a researcher or autonomous system astray or greatly delay reaching the researcher's goals, e.g., identifying a novel, optimal material. The materials community needs a diverse set of data on which to benchmark active learning schemes. These datasets would serve as surrogates for the challenges of materials exploration and optimization. From these benchmark studies, one can then identify the pros and cons of applying each scheme to different materials challenges.

In this work, we present a reference materials dataset for benchmarking common off-the-shelf science agnostic BO schemes along with one scientific ML[14] BO scheme for the task of materials optimization. Here, scientific ML refers to ML algorithms with built in prior physical knowledge. The dataset is from a fully-characterized thin film library of the Fe-Co-Ni[15] system where structural properties and magnetic properties were mapped across the entire ternary (Figure 2). Using this complete data, one can build a highly accurate composition-structure-property model, which can be used for the purpose of evaluating the efficacy of different active learning schemes through simulating closed-loop experiments (see Figure 1). The active learning scheme identifies a material composition spot on the library to investigate; pertinent data is then collected from the model to simulate experiment execution; machine learning is then used to analyze the collected data, and the process repeats. Here, we use this database to present the first benchmark of active learning schemes for materials optimization and discovery in general and for screening optimal materials in combinatorial libraries. A previous study benchmarked active learning schemes for composition-phase determination on the Fe-Ga-Pd dataset[16]. These benchmark datasets along with others are available at the Resource for Materials Informatics website[17].

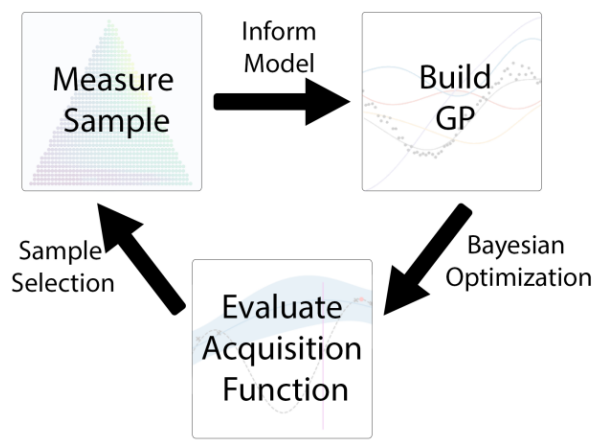

Figure 1. Representation of the active learning (AL) pipeline using Gaussian process as the machine learning surrogate function and Bayesian optimization for materials optimization. The AL framework guides subsequent experiments to hone-in on the target materials.

## DISCUSSION

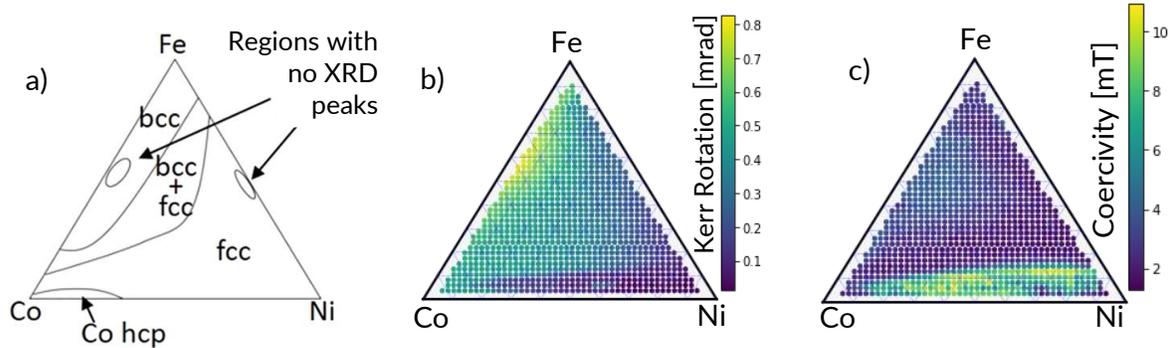

Figure 2. The Fe-Co-Ni library dataset[15]. a) The composition-structural phase map for the library. This map was determined through expert analysis of collected X-ray diffraction data. The two materials optimization challenges have been chosen for their varying complexity. b) The Kerr rotation (measured in milliradians) is the least complex dataset. As seen visually, the function is dominated by a large gradually increasing peak with a maximum located along the Fe-Co binary axis. c) The magnetic coercivity (measured in milli-Tesla) represents a high complexity system with the global maximum and many local maxima located in a small region near the Ni-Co binary axis and one broad small peak located along the Fe-Co binary axis.

A thin film Fe-Co-Ni composition library dataset[15] is used to investigate materials optimization performance of varying active learning schemes. Expert-based data cleaning was performed to simplify benchmark use and a description of these preprocessing steps is given in the Supplemental Information. Sample materials span the full ternary composition space of Fe-Co-Ni, with 921 compositions/samples in total. Structure data is collected for each sample, providing information of the ternary composition-structural phase diagram (Fig 2a), i.e., how lattice structure varies across the composition space. Two materials properties were measured for each sample - Kerr rotation (Fig 2b) and magnetic coercivity (Fig 2c). The Kerr rotation is a measure of magnetization of the material, while the coercivity is a measure of the magnetic hardness[18]. Here the composition dependence of the Kerr rotation is shown to be the less complex of the two, smoothly varying with a broad peak near the binary of $Fe_{0.4}Ni_{0.6}$. Coercivity shows a broad peak at a similar composition which is overwhelmed by a higher, complex and more "granular" coercivity response near the Co-Ni binary with many local optima.

The list of off-the-shelf active learning acquisition functions[13] benchmarked is presented in Table 1. The off-the-shelf acquisition functions are paired with a Gaussian Process for the prediction model. Here, $\mu$, $\sigma^2$, and $\Sigma$ are the GP mean, variance, and covariance, respectively. $f(x^+)$ is the maximum found in the previous iterations; $n$ is the current iteration number; $D$ is the number of materials to search over; $\lambda$ is a predefined constant, here set to 0.1; the constants $\sqrt{\ln(Dn^2\pi^2)/3\lambda}$ and $\sqrt{\ln(2n+1)/8}$ balance exploration typified by maximizing $\sigma^2$ and exploitation typified by maximizing $\mu$; $N$ is the probability density function for the normal distribution, and $\Phi$ and $\phi$ are the cumulative distribution function and the probability distribution function for the standard normal distribution $N(\mu = 0, \sigma = 0)$ respectively. The off-the-shelf-acquisition functions are paired with a GP using an isometric radial basis function and Gaussian likelihood. These active learning schemes are compared to a recent physics-informed active learning scheme named Closed-loop Autonomous Materials Exploration and Optimization (CAMEO)[4]. CAMEO exploits a fundamental rule in materials science, that a material system's structure, or phase, is predictive of its functional properties. As such, CAMEO first seeks to maximize knowledge of the target material system's phase diagram. It then uses phase boundaries to segment the search space and guide the search for materials optimization.

The materials optimization process is initialized by selecting a material with uniform probability. The GP is then used to predict the material property of interest for all materials including those without functional property data. The acquisition function of interest is used to select the next material to investigate. This loop is continued exhaustively for the remainder of the unevaluated data points (materials cannot be selected twice in this implementation) with the same acquisition function. For each acquisition function the process is repeated 100 times. Performance is computed using minimum regret, defined as:

$$minimum\ regret = \max(sampled) - \max(global) \quad \text{Eq. (1)}$$

The mean and mean confidence intervals for the minimum regret are computed over the 100 runs, and the mean is plotted in Figures 3 and 4 for the Kerr rotation and coercivity respectively. Figures with both mean and the confidence intervals plotted are available in the Supplemental Information.

Table 1. Acquisition functions

| Name | Acquisition Function | Description |
| --- | --- | --- |
| Exploration | $\text{argmax}_x[\sigma^2]$ | Point where model has max uncertainty |
| Exploitation | $\text{argmax}_x[\mu]$ | Point with predicted max value |
| Upper Confidence Bound (UCB) | $\text{argmax}_x[\mu + \sigma\sqrt{\ln(Dn^2\pi^2)/3\lambda}]$ | Balance of exploration and exploitation with iteration $n$ dependent balance ratio |
| Add-GP-UCB[19] | $\text{argmax}_x[\mu + \sigma\sqrt{\ln(2n+1)/8}]$ | Balance of exploration and exploitation with iteration dependent balance ratio |
| Thompson Sampling | $\text{argmax}_x[f \sim N(\mu, \Sigma)]$ | Sample function $f$ from Gaussian distribution with model's predicted mean and covariance. Then identify the point with max value. |
| Expected Improvement | $\text{argmax}_x[(\mu - f(x^+) - \xi)\Phi(Z) + \sigma\phi(Z)]$ $Z = (\mu - f(x^+) - \xi)/\sigma$ | Identify point expected to have maximal improvement over past identified maximum |
| Random Sampling | $x^* \sim Unif(X)$ | Sample point at random |

For the lower complexity Fe-Co-Ni Kerr rotation dataset, locating the one dominant peak can be achieved with a simple estimated gradient ascent method[7]. As a result, greedier acquisition functions that focus on exploitation have better performance. Nevertheless, all acquisition functions reach within .1% deviation from the maximum within 5% of the 921 data points. Add-GP-UCB[19] reaches the goal within approximately 11 samples; about double the speed of its next competitor. In second place is the physics-informed CAMEO algorithm which must expend initial iterations to identify the phase diagram. This forced exploration puts CAMEO at a disadvantage when more aggressive exploration provides better performance. All acquisition functions perform significantly better than random sampling due to the simplicity of the target function. If the data is known beforehand to have gradual changes in intensity and one dominant peak, an exploitation-focused acquisition algorithm is preferable.

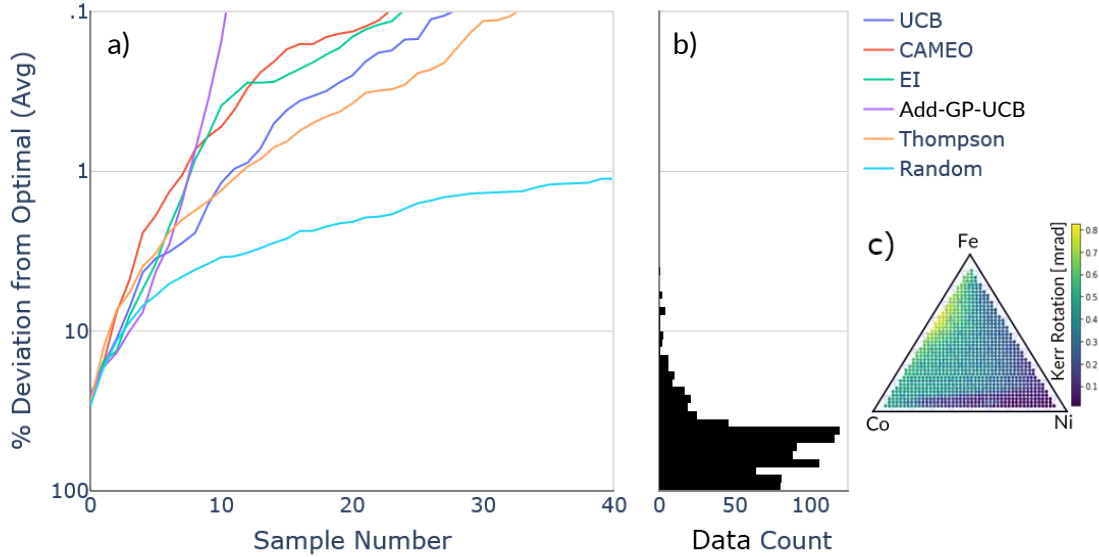

Figure 2. a) Benchmarking performance of acquisition functions on Fe-Co-Ni Kerr Rotation with reference to random selection. b) A histogram displaying the counts of data at different % deviations from the optimal. c) The Fe-Co-Ni Kerr rotation dataset replotted.

The high-complexity material dataset of Fe-Co-Ni magnetic coercivity has a large broad maximum near the Fe-Co binary and high roughness/variation with many local maxima in the proximity of the global maximum, all of which can serve to distract an acquisition function. The high roughness near the global maximum results in a small indicative composition region, making the maximum difficult to find. The majority of acquisition functions perform poorly, easily becoming stuck in local maxima. Expected Improvement performs extremely well in comparison to other common acquisition functions, and is more exploratory than the other methods, while still outperforming random. While the physics-informed CAMEO algorithm is capable of narrowing the search space to those phase regions that promise to hold a maximum, the high roughness of the target phase regions still manages to distract CAMEO from the global maximum. This is likely due to the use of the UCB algorithm once CAMEO switches from phase mapping to materials optimization. The number of samples required to reach .1 % from the optimum is larger than that of the simpler Kerr rotation challenge.

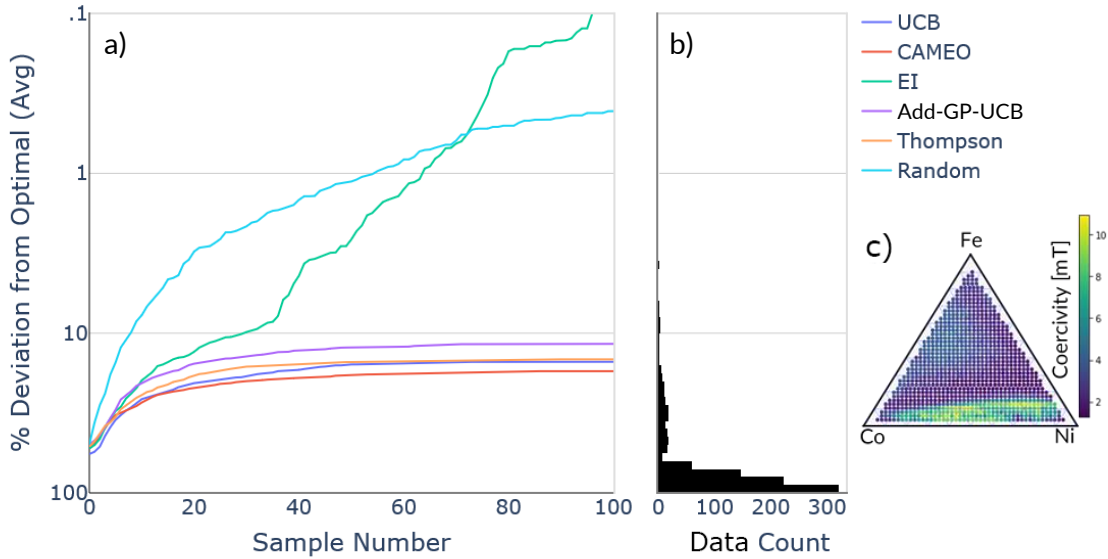

Figure 3. a) Benchmarking performance of acquisition functions on Fe-Co-Ni magnetic coercivity dataset with reference to random selection. b) Adjacent to the performance graph is a histogram displaying the counts of data at different % deviations from the optimal. c) The dataset is complex, with a broad, minor peak near the Fe-Co binary and high roughness with many local maxima near the global maximum close to the Co-Ni binary.

     The Fe-Co-Ni Kerr dataset of composition, X-ray diffraction, Kerr rotation, and magnetic coercivity provides some interesting insights into active learning use for real-world tasks materials optimization. When a smooth, simple landscape is identified, schemes with greater focus on exploitation succeed. The alternative is also true with exploration being preferable with highly complex landscapes. This suggests an active learning scheme that is iteration dependent. At each iteration one can compare prediction accuracy and quantify landscape complexity to determine the preferred acquisition function for the next iteration, shifting the balance between exploration and exploitation. Nevertheless, expected improvement is the overall winner, performing well for either task, despite the incredible complexity of the coercivity data.

     The physics-informed CAMEO method lags in performance when the challenge is simple, due to its forced steps of iteration – it must first converge on a phase diagram before seeking an optimal material. It is also able to narrow the search space with a highly complex landscape, but the currently choice of UCB for optimization within the target phase region performs poorly and should potentially be replaced by an alternative method such as expected improvement. The authors hope that the Fe-Co-Ni and Fe-Ga-Pd[4] datasets can spur interest in benchmarking and developing novel active learning schemes for real-world challenges of materials exploration and optimization.

**Acknowledgement:**
This work was funded by NIST Cooperative Agreement 70NANB17H301 and an Office of Naval Research MURI through grant #N00014-17-1-2661.

\* **NIST Disclaimer:** Certain commercial equipment, instruments, or materials are identified in this report in order to specify the experimental procedure adequately. Such identification is not intended to imply recommendation or endorsement by the National Institute of Standards and

# Supplemental Information

**Fe-Co-Ni Figures with Mean and Confidence Intervals**

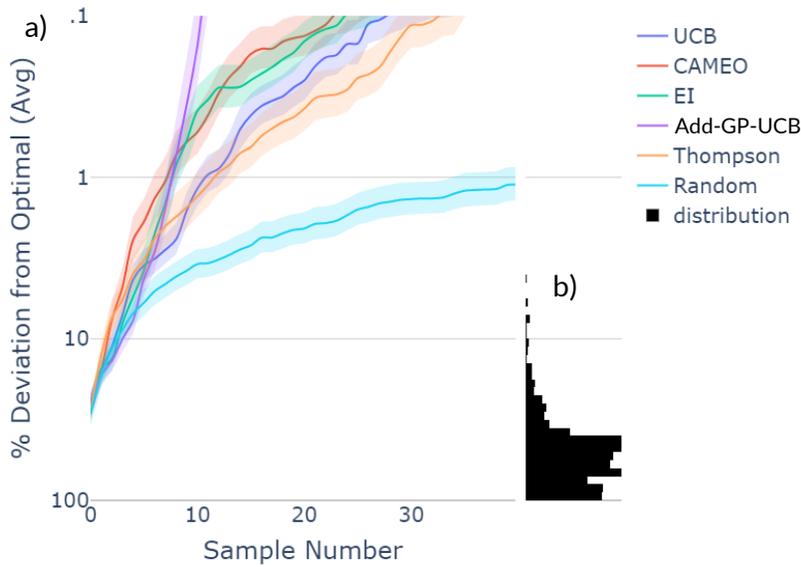

Figure 2. a) Benchmarking performance of acquisition functions on Fe-Co-Ni Kerr Rotation with reference to random selection. Shaded regions are 95 % confidence intervals for the mean. b) A histogram displaying the counts of data at different % deviations from the optimal.

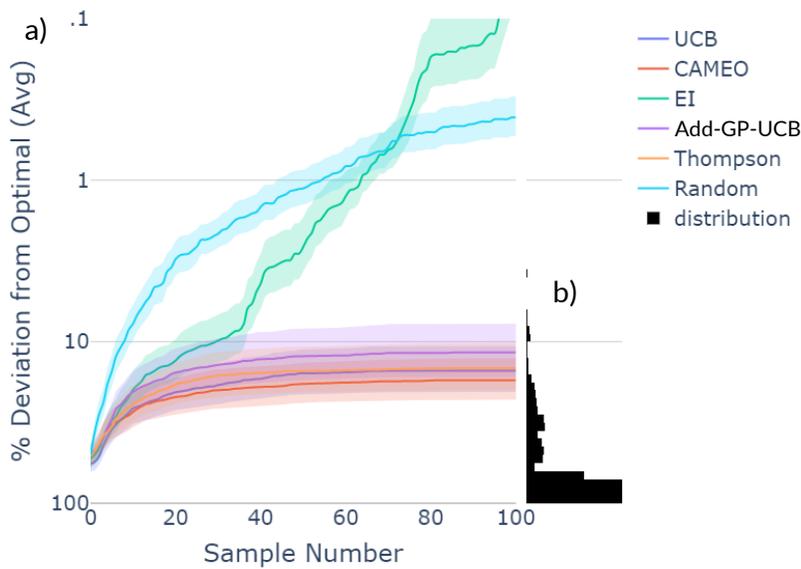

Figure 3. a) Benchmarking performance of acquisition functions on Fe-Co-Ni magnetic coercivity dataset with reference to random selection. Shaded regions are 95 % confidence intervals for the mean. b) Adjacent to the performance graph is a histogram displaying the counts of data at different % deviations from the optimal.

**Confidence Interval**

The 95 % confidence interval was computed for the variable of interest over 100 experiments at the given iteration with:

$$CI_{95} = \left(\frac{\sigma}{\sqrt{n}}\right) F^{-1}(p, v) \qquad (12)$$

Where $F^{-1}$ is the inverse of the Student's t cumulative distribution function, $\sigma$ is the standard deviation, $n = 100$ is the number of experiments, $p = \{2.5\%, 97.5\%\}$, and $v = 99$ is the degrees of freedom.

**Data Preprocessing**

The initial Fe-Co-Ni dataset contains 1252 data points compared to the 921 data points presented here. The data cleaning process follows these steps:
- Using composition measurements, samples associated with substrate measurements (composition percentages for Fe, Co, and Ni are zero) are removed, giving 1128 samples.
- Samples along the edges of the composition spread exhibit edge effects, e.g., low film thickness and the resulting poor representation of Fe-Co-Ni behavior. These samples are removed, giving 921 samples.

The data preprocessing steps are available as a Matlab* script.